\documentstyle[amssymb]{elsart}
\def\openone{\leavevmode\hbox{\small1\kern-3.8pt\normalsize1}}%
\def\hilb#1{\hbox{${\cal H}_{#1}$}}%
\def\<{\langle} \def\>{\rangle}%
\def\ket#1{\vert {#1} \>}%
\def\kket#1{\vert {#1} \>\!\>}%
\def\kketp#1#2{\vert {#1} \>\!\>_{#2}}%
\def\bra#1{\< {#1} \vert}%
\def\bbra#1{\<\!\< {#1} \vert}%
\def\bbrap#1#2{\,_{#2}\<\!\< {#1} \vert}%
\def\prj#1{\ket{#1}\bra{#1}}%
\def\pprjp#1#2{\kketp{#1}{#2}\bbrap{#1}{#2}}
\def\pprj#1{\kket{#1}\bbra{#1}}
\def\eex#1#2{\bbra{#1}{#2}\>\!\>}

\def\grp{\hbox{$\mathbf G$} }
\def\grpZZ{{\mathbb Z}_N\times{\mathbb Z}_N}

\def\cmplx{{\mathbb C}}
\def\beps{{\cal E}}
\def\bid{{\cal I}}
\def\btau{\hbox{\Large $\tau$}}

\begin{document}
\begin{frontmatter}
\title{Bell Measurements and Observables} 
\author{G. M. D'Ariano, P. Lo Presti and M. F. Sacchi}
\address{Theoretical Quantum
Optics Group\\ Universit\`a degli Studi di Pavia and INFM Unit\`a di
Pavia\\ via A. Bassi 6, I-27100 Pavia, Italy} 
\begin{abstract}
A general matrix approach to study entangled states is presented,
based on operator completeness relations. Bases of unitary operators
are considered, with focus on irreducible representations of groups.
Bell measurements for teleportation are considered, and robustness of
teleportation to various kinds of non idealities is shown.
\end{abstract}
\end{frontmatter}
\section{Introduction}
Quantum mechanics builds up systems from subsystems in a fascinating
way, through the tensor product, that allows one to set up the so
called entangled states. These are states of the whole system that do
not correspond to any state of the subsystems taken separately. This
peculiar aspect of quantum world stands at the foundations of all the
recent developments of quantum information theory, such as dense
coding, teleportation, quantum computation, quantum cryptography, and so
on \cite{pope}. These theoretical results have recently entered the realm of
experimental physics \cite{exper}.

Analogously to what happens for states, also quantum measurements on
composite systems can be entangled when they are non local, namely
they cannot be considered as a measurement jointly performed on the
subsystems.  In the general framework of positive operator valued
measures (POVM), entangled measurements correspond to non factorizable
POVM's. The so called ``Bell measurements'' are the most relevant
example \cite{BBCJPW}, corresponding to maximally entangled POVM's.
Entanglement and Bell measurements are the basic ingredients of
quantum teleportation.

In this letter, we present a matrix approach to address
bipartite-system pure states along with general operator-completeness
relations. These allow us to write the most general Bell-like POVM in
compact form. Bases of unitary operators are considered, with focus on
irreducible representation of groups. The canonical role of the groups
$\grpZZ$ and {\it Weyl-Heisenberg} is analyzed. We conclude with a study of robustness
of teleportation to different kinds of non idealities.

\section{Operator ``basis''}

Consider a set of linear operators $\{B(\lambda),\;\lambda\in\Sigma,\;\Sigma\;
\hbox{Borel space}\}$ on a finite dimensional Hilbert space
\hilb{}. This set is a ``spanning set'' for the operator space
 if it satisfies one of the following equivalent statements:
\begin{itemize}
\item[1.] Completeness relation:
\begin{eqnarray}
&&Tr[B^\dagger(\lambda)B(\lambda')]=\Delta(\lambda,\lambda')\;
,\quad\hbox{where}
\quad\int_\Sigma d\lambda\,\Delta(\lambda,\lambda')\,B(\lambda)=B(\lambda')\;,
\nonumber 
\\
&&\hbox{and}\qquad Tr[B^\dagger(\lambda)A]=0\quad\forall \lambda\qquad
\Leftrightarrow\qquad A=0\;.\label{completeness-b}
\end{eqnarray}
\item[2.] For any linear operator $A$ on \hilb{},
\begin{equation}
\int_\Sigma d\lambda \, Tr[B^\dagger(\lambda)A]\,B(\lambda)=A\;.
\label{A on operator basis}
\end{equation}
\item[3.] Chosen any orthonormal basis $\{\ket i\}$ for \hilb{},
\begin{equation}
\int_\Sigma d\lambda \, \bra{n}B^\dagger(\lambda)\ket{m}\bra{l}B(\lambda)
\ket{k}=\delta_{nk}\delta_{ml}\;.
\label{orthogonality}
\end{equation}
\item[4.] For any linear operator $A$ on \hilb{},
\begin{equation}
\int_\Sigma d\lambda\,B^\dagger(\lambda)\,A\,B(\lambda)=Tr[A]\,\openone\;.
\label{depolarizing}
\end{equation}
\end{itemize}
{\it Proof of $1\Leftrightarrow 2$:}\\ To prove $(\Rightarrow)$ we
define $O=\int Tr[B^\dagger(\lambda)A]B(\lambda)d\lambda-A$. Then we
evaluate the following trace
\begin{equation} 
Tr[B^\dagger(\lambda')O]= \int
Tr[B^\dagger(\lambda')B(\lambda)]\,Tr[B^\dagger(\lambda)A]\,d\lambda-
Tr[B^\dagger(\lambda')A]=0\;,
\end{equation}
where integration has been carried out by means of the first line of
Eq.  (\ref{completeness-b}); the second line of
Eq. (\ref{completeness-b}) completes the proof.  Converse implication:
the first line of Eq.  (\ref{completeness-b}) follows immediately by
replacing $A$ with $B(\lambda')$ in Eq. (\ref{A on operator basis}),
whereas the second part is a direct consequence of Eq. (\ref{A on
operator basis}).
\\
{\it Proof of $2\Leftrightarrow 3$:}\\ The proof of $(\Rightarrow)$ is
immediate by substituting $A$ with $\ket{m}\bra{n}$ in Eq. (\ref{A on
operator basis}) and taking the matrix element between $\bra{l}$ and
$\ket{k}$. The converse is also straightforward: multiply both members
of Eq. (\ref{orthogonality}) by $\bra{m}A\ket{n}\ket{l}\bra{k}$ and
take the sum over all indices $k$, $l$, $m$, $n$.
\\
{\it Proof of $3\Leftrightarrow 4$:}\\ The direct implication is
derived multiplying both members of Eq. (\ref{orthogonality}) by
$\bra{m}A\ket{l}\ket{n}\bra{k}$, and summing the result over all the
indices $k$, $l$, $m$, $n$.  To prove the converse, let
$A=\ket{m}\bra{l}$ in Eq. (\ref{depolarizing}) and take the matrix
element between $\bra{n}$ and $\ket{k}$.

Note that Eq. (\ref{A on operator basis}) is exactly the linear
decomposition of the operator $A$ on a set of operators $\{B(\lambda)\}$
induced by the scalar product $(B,A)=Tr[B^\dagger A]$.  For infinite
dimensional Hilbert spaces the previous relations have meaning for
Hilbert-Schmidt operators. However, they still hold for all linear
operators in a distribution sense.

\section{General representation of bipartite-system pure states}
Chosen two orthonormal bases $\{\ket{i}_1\}$ and $\{\ket{j}_2\}$ for
the Hilbert spaces \hilb1 and \hilb2 respectively, any vector
$\kket{\psi}\in\hilb1\otimes\hilb2$ can be written as
\begin{equation}
\kket{\psi}=\sum_{ij}c_{ij}\ket{i}_1\ket{j}_2\doteq\kket{C}\;.
\label{kket}
\end{equation}
Eq. (\ref{kket}) introduces a notation that exploits the
correspondence between vectors in $\hilb1\otimes\hilb2$ and
$N\times M$ matrices, where N and M are the dimensions of \hilb1 and \hilb2, 
respectively (cfr. Ref. \cite{royer}).
\\ The following relations are an immediate consequence of Eq. (\ref{kket})
\begin{eqnarray}
\hskip 2cm
&A\otimes B\kket{C}=\kket{ACB^T}\;,\;\;\;\eex{A}{B}=Tr[A^\dagger
B]\;,\nonumber& \\
&Tr_2\bigg[\kketp{A}{12}\bbrap{B}{12}\bigg]=(AB^\dagger)^{(1)}\;,\nonumber&\\
&Tr_1\bigg[\kketp{A}{12}\bbrap{B}{12}\bigg]=(A^TB^*)^{(2)}\;.&
\label{kket formulas}
\end{eqnarray}
Notice that the definition of the matrix $C$ in Eq. (\ref{kket}) is
base-dependent, hence the transposition and conjugation in Eqs.
(\ref{kket formulas}) are referred to the same fixed basis.  These
relations are very useful for derivations and to express the
results in an index-free compact form.

In the following we will focus our attention on bipartite systems
whose Hilbert space is $\hilb{}\otimes\hilb{}$, with $N=dim(\hilb{})$. As
an application of the formalism just introduced, we give a direct
proof of the existence of the Schmidt decomposition for a pure state
of a bipartite system. Using a polar decomposition $A=V\sqrt{A^\dagger
A}$, with $V$ unitary, which holds for any matrix $A$
 \cite{polar decomposition}, we choose a unitary
operator $U$ so that $UA^\dagger A U^\dagger$ is diagonal, then we can
write
\begin{equation}
\kket{A}=\kket{V\sqrt{A^\dagger A}}=
VU^\dagger\otimes U^T \kket{U\sqrt{A^\dagger A}U^\dagger}=
\sum_{i}\sqrt{\lambda_i}\ket{i}_1'\ket{i}_2''\;, 
\label{Schmidt}
\end{equation}
where $\ket i_1'=VU^\dagger\ket i_1$, $\ket i_2''=U^T\ket i_2$ and
$\lambda_i$ is the eigenvalue of $A^\dagger A$ with eigenvector $\ket{i}$.

Using Eq. (\ref{kket}), it is straightforward to characterize
maximally entangled states. These are defined as the states $\kket{A}$
whose partial trace on each of the two subsystems is proportional to
identity; namely
\begin{equation}
Tr_1\bigg[\pprj{A}\bigg]=A^TA^*=\frac1N \openone\quad
\hbox{and}\quad
Tr_2\bigg[\pprj{A}\bigg]=AA^\dagger=\frac1N\openone\;,
\end{equation}
hence maximally entangled states are of the form
\begin{equation}
\kket A =\frac1{\sqrt{N}} \kket U
\;,\label{max ent}
\end{equation}
with $U$ unitary. 
Two maximally entangled states are always connected by means of a
local unitary transformation. In fact
\begin{equation}
\kket U =UV^\dagger\otimes\openone\kket V\;.
\end{equation}
Given a spanning set $\{B(\lambda)\}$, the set of vectors
$\{\kket{B(\lambda)}\}$ spans $\hilb{}\otimes\hilb{}$ in the sense
that
\begin{equation}
 \kket{A}=\int d\lambda\, Tr[B^\dagger(\lambda)A]\,\;\kket{B(\lambda)}\;.
\end{equation}
Moreover one has 
\begin{eqnarray}
 \int d\lambda\,\pprj{B(\lambda)}&=&\int d\lambda\,B(\lambda)\otimes\openone
   \;\pprj{\openone}\;B^\dagger(\lambda)\otimes\openone=\nonumber\\
   &=&Tr_1[\pprj \openone]=\openone\;, 
\label{B POVM}
\end{eqnarray}
hence the projectors on $\kket{B(\lambda)}$ provide a resolution of
the identity and a POVM.
\\By explicit evaluation of the matrix elements, one can easily verify
the following useful formulas
\begin{equation}
\int d\lambda\,B(\lambda)\otimes B^*(\lambda)=\pprj{\openone}\;,
\label{std ent}
\end{equation}
\begin{equation}
\int d\lambda\,B(\lambda)\otimes B^\dagger(\lambda)\kket{A}=\kket{A^T}\;,
\label{transposer}
\end{equation}
which are directly equivalent to Eq. (\ref{orthogonality}) of statement 3.
\section{Bell Measurements}
A Bell measurement is a POVM whose elements are projectors on
maximally entangled states. Referring to Eqs. (\ref{max ent}) and
(\ref{B POVM}) we argue that any POVM of this kind corresponds to a
spanning set whose elements are proportional to unitary operators
\begin{equation}
\Pi(d\lambda)=\pprj{\tilde U(\lambda)}d\lambda\;,
\label{Bell POVM}
\end{equation}
where $\tilde U(\lambda)$ is a basis with $\tilde
U(\lambda)=\alpha(\lambda)U(\lambda)$, $U(\lambda)$ unitary, and
$\alpha(\lambda)$ c-number.

As  proved in Ref. \cite{NielsenCaves}, Bell measurements
are the only projector valued POVM's capable of teleportation in the
case of pure preparation of the shared resource, which turns out to be
necessarily in a maximally entangled state.
\\In the following we give a brief description of this kind of
teleportation scheme.  The Hilbert space \hilb 1 is prepared in an
unknown state $\rho^{(1)}$, whereas $\hilb 2\otimes\hilb3$ is in the
maximally entangled state $\frac{1}{\sqrt{N}}\kketp{\openone}{23}$
(\hilb{1,2,3} have the same dimension). Upon performing the
measurement described by the POVM (\ref{Bell POVM}) on $\hilb 1\otimes
\hilb 2$, the (unnormalized) state  on \hilb 3 conditioned by the outcome
$\lambda$  will be
\begin{eqnarray} 
\tilde \varrho_\lambda^{(3)}
&=&Tr_{12}\bigg[\rho^{(1)}\otimes\pprjp{\openone}{23}\;\;
\pprjp{\tilde U_\lambda}{12}\otimes\openone_3\bigg]=\nonumber\\
&=& Tr_{12}\bigg[\rho^{(1)}\otimes\pprjp{\openone}{23}\;\; \tilde
U_\lambda^{(1)}\otimes\openone^{(23)}\times\nonumber\\&\times&
\pprjp{\openone}{12}\otimes\openone^{(3)}\;\; \tilde
U_\lambda^{\dagger(1)}\otimes\openone^{(23)}\bigg]=\nonumber\\ &=&
\bbrap{\openone}{12}\kketp{\openone}{23}\;\tilde
U_\lambda^{\dagger(1)}\rho^{(1)}\tilde U_\lambda^{(1)}\;
\bbrap{\openone}{23}\kketp{\openone}{12}=\tilde
U_\lambda^{\dagger(3)}\rho^{(3)}\tilde U_\lambda^{(3)}\;.
\label{state reduction}
\end{eqnarray}
The normalized state writes
\begin{equation}
\varrho_\lambda=U^\dagger(\lambda)\,\rho\,U(\lambda)\;,
\label{norm state red}
\end{equation}
and the teleportation can be completed upon applying the unitary
transformation $U(\lambda)$ on the state (\ref{norm state red}).  If
the shared entangled resource is prepared in another maximally
entangled state, i.e. $\frac{1}{\sqrt{N}}\kketp{V}{23}$ with V
unitary, it is enough to substitute $U(\lambda)$ with
$U(\lambda)V^*$.\\
Notice that the product $\bbrap{\openone}{12}\kketp{\openone}{23}$
that appears in Eq. (\ref{state reduction}) corresponds to the
transfer operator $\btau_{31}$ of Ref. \cite{BraunD'ArianoMilbSacchi},
which for any vector $\ket{\psi}_1$ of \hilb1 satisfies the relation
\begin{equation}
\btau_{31}\ket{\psi}_1=\ket{\psi}_3\;.
\label{transfer}
\end{equation}
If the set $\{\tilde U(\lambda)\}$ is an orthonormal operator basis
(Dirac-like orthonormality relations are allowed in the case of
infinite dimensional spaces), it is possible to write the class of
Bell observables, i.e. the self-adjoint operators that one has to
measure in order to realize the Bell measurement. The Bell observables
can be written as follows
\begin{equation}
O=\int f(\lambda)\,\pprj{\tilde U(\lambda)}\,d\lambda\;,
\label{Bell observable}
\end{equation}
where $f(\lambda)$ must be an injective function (i.e. O is non
degenerate) in order to guarantee a univocal correspondence between
the read eigenvalue $f(\lambda)$ and the unitary operator $U(\lambda)$ of
Eq. (\ref{norm state red}) that completes the teleportation scheme.

\subsection{The role of group representations}
Unitary irreducible representations (UIR) of groups provide a method
to generate a spanning set of unitary operators in the sense of statements
(\ref{completeness-b}--\ref{depolarizing}). In fact, if
$\{U_g,\;g\in\grp\}$ are the elements of a projective UIR 
of the group $\grp $, from the first Schur's lemma it
follows that
\begin{equation}
\int_G dg\,U_g \,A\, U^\dagger_g=Tr[A]\openone\;,
\label{trace formula}
\end{equation}
where $dg$ is a (suitably normalized) group invariant measure on
\grp. Recalling Eq. (\ref{Bell POVM}), it follows that the POVM
\begin{equation}
\Pi(dg)=\pprj{U_g}dg
\end{equation}
describes a Bell measurement.\\ For example, as noticed in
Ref. \cite{BraunD'ArianoMilbSacchi}, the N-dimensional UIR of the
group $\grpZZ$ whose elements are
\begin{equation}
U(m,n)=\sum_ke^{2\pi ikm/N}\ket{k}\bra{k\oplus n}\;,
\label{ZZ representation}
\end{equation}
generates the Bell measurement corresponding to the teleportation
scheme of Ref. \cite{BBCJPW}.

As an example for the infinite dimensional case, consider
the displacement operators of an electromagnetic
field mode $a$ ($[a, a^\dagger]=1$)
\begin{equation}
D(z)=\exp(za^\dagger- z^* a)\;,\qquad z \in \cmplx\;.
\label{displacement}
\end{equation}
Such operators are the elements of a projective UIR representation of
the Weyl-Heisenberg group $WH$, and generate the Bell measurement
corresponding to the Braunstein-Kimble teleportation scheme of
Ref. \cite{BraunsteinKimble}.

For $\grpZZ$ the class of Bell observables defined by Eq. (\ref{Bell
observable}) is given by
\begin{eqnarray}
O&=&\sum_g f(g)\pprj{U_g}
 =\sum_g f(g)\,U_g\otimes\openone\;\sum_{g'}U_{g'}\otimes U_{g'}^*\;U_g^\dagger
    \otimes\openone =\nonumber\\
 &=&\sum_{m,n}\sum_{m',n'}f(m,n)\,e^{\frac{2\pi i}{N}(nm'-mn')}\,
     U(m',n')\otimes U^*(m',n')=\nonumber\\
 &=&\sum_g \tilde f(g)\,U_g^{(1)}\otimes U^{(2)*}_g\;, 
\end{eqnarray}
where we used Eq. (\ref{std ent}) along with the relation
\begin{equation}
U(m,n)U(m',n')U^\dagger(m,n)=e^{\frac{2\pi i}{N}(nm'-mn')} U(m',n')\;,
\end{equation}
and we introduced the Fourier transform $\tilde f$ over the group
\begin{equation}
\tilde f(m,n)=\sum_{m',n'}e^{\frac{2\pi i}{N}(nm'-mn')}f(m',n')\;.
\end{equation}

By applying Eq. (\ref{std ent}), the analogous relation for $WH$ reads
as follows
\begin{eqnarray}
O&=&\int_\cmplx d^2z\,f(z)\,\pprj{D(z)}=\nonumber\\
&=&\int_\cmplx d^2z\,f(z)D(z)\otimes\openone \int_\cmplx \frac{d^2\alpha}{\pi}
  D(\alpha)\otimes D(\alpha^*)\;\; D(z)^\dagger\otimes\openone=\nonumber\\
&=&\int_\cmplx d^2\alpha \int_\cmplx \frac{d^2z}{\pi}\,f(z)\,
  e^{\alpha z^*-\alpha^* z}D(\alpha)\otimes D(\alpha^*)=\nonumber\\
&=&\int_\cmplx d^2\alpha\,\tilde f(\alpha)\,D(\alpha)\otimes D(\alpha^*)\;.
\end{eqnarray}
However, in this case, one can derive a more explicit expression for
the Bell observables. In fact, from Eq. (\ref{std ent}) one has
\begin{eqnarray}
\pprjp{\openone}{12}
&=&\int_\cmplx \frac{d^2\beta}{\pi}D_1(\beta)\otimes D_2(\beta^*)=\nonumber\\
&=&\int_\cmplx \frac{d^2\beta}{\pi}\exp\bigg[(\beta a_1^\dagger-\beta^* a_1)
   +(\beta^* a_2^\dagger- \beta a_2)\bigg]=\nonumber\\
&=&\int_\cmplx \frac{d^2\beta}{\pi}
   \exp\bigg[\beta Z^\dagger_{12} -\beta^* Z_{12}\bigg]
   \doteq\pi\delta^{(2)}(Z_{12})\;,
\end{eqnarray}
with $Z_{12}=a_1-a_2^\dagger$.
Using the relation $D_a(z)\,a\,D^\dagger_a(z)=a-z$, one obtains
\begin{equation}
\frac1{\pi}\pprjp{D(z)}{12}=\delta^{(2)}(Z_{12}-z)\;,
\end{equation}
and finally
\begin{equation}
O =\int_\cmplx d^2z f(z)\frac1{\pi}\pprjp{D(z)}{12}=  f(Z_{12})\;.
\end{equation}
Hence, in order to realize the Bell measurement generated by $WH$, we
have to measure an injective function of the operator $Z_{12}$, or
simply $Z_{12}$ itself. This measurement can be easily performed by
unconventional heterodyne detection (cfr. Ref. \cite{unc}). 

\section{Robustness of ``pure'' teleportation}

``Pure'' teleportation schemes rely on projector valued POVM's and
pure preparations for the shared resource. As proved in
Ref. \cite{NielsenCaves}, this kind of teleportation works properly if
and only if the elements of the POVM are proportional to projectors on
maximally entangled states and the resource itself is maximally
entangled.  However, for practical purposes, one is interested in the
evaluation of the robustness of this kind of schemes to non
ideality.\\ 
Looking at Eq. (\ref{state reduction}), it is evident that
the state on $\hilb 3$ conditioned by the measurement is a continuous
function of the shared resource preparation and of the element
of the POVM related to the outcome.  Since the teleported state is
again a continuous function of this conditioned state and of the
``adjusting'' unitary transformation, we conclude that teleportation is
robust to non ideal entanglement preparation, non ideal measurement,
and non ideal adjusting transformation.

Let's suppose that before the measurement the maximally entangled
resource evolves according to a trace preserving CP map $\beps$ owing
to some kind of noise. In the following, we will simply evaluate the
state on $\hilb 3$ after the measurement in presence of such a noise.
\\By means of the Kraus's decomposition \cite{Kraus} of $\beps$, the
noisy state of $\hilb 2\otimes\hilb 3$ can be written as
\begin{equation}
\rho_{23}=\beps\bigg(\frac1N\pprjp{\openone}{23}\bigg)=
\sum_\mu A_\mu^{(23)}\; \frac1N\pprjp{\openone}{23}\;A^{(23)\dagger}_\mu\;,
\label{noise map}
\end{equation}
where $A_\mu$ are operators on $\hilb 2\otimes \hilb 3$ satisfying
$\sum_\mu A^\dagger_\mu A_\mu =\openone$.
\\For any (generally non local) operator A acting 
on $\hilb{}\otimes\hilb{}$ one has
\begin{equation}
A\kket{\openone}=\kket{\hat A^T}=\openone\otimes \hat
A\kket{\openone}\;,
\label{nonloca to local}
\end{equation}
where $(\hat A)_{i,j}=\sum_l \langle i |\langle j | A |l \rangle |l\rangle
  $. Therefore it is possible to write
\begin{eqnarray}
\beps\bigg(\frac1N\pprjp{\openone}{23}\bigg)&=&
\sum_\mu A_\mu^{(23)}\;\frac1N\pprjp{\openone}{23}\;A^{(23)\dagger}_\mu
= \nonumber\\
&=&\sum_\mu\openone_2\otimes\hat A^{(3)}_\mu \;\frac1N\pprjp{\openone}{23}\;
\openone_2\otimes\hat A^{(3)\dagger}_\mu=\nonumber\\
&=&\bid^{(2)}\otimes\hat\beps^{(3)}\bigg(\frac1N\pprjp{\openone}{23}\bigg)\;,
\label{local map}
\end{eqnarray}
where $\hat \beps$ is the map whose Kraus's decomposition is given by
the operators $\hat A_\mu$. This last equation shows how the action of
any CP map on a maximally entangled state of a bipartite system can be
written as the result of the application of a local CP map.\\
Recalling Eq. (\ref{state reduction}), it results that the local CP
map $\hat\beps^{(3)}$, which describes noise, commutes with all other
maps and with the partial trace, so that the unnormalized conditioned
state after the measurement, in presence of such a noise, can be
simply written as
\begin{equation}
\tilde\varrho_\lambda^{(3)}=\hat \beps^{(3)}(\;U(\lambda)^\dagger \,
              \rho \, U(\lambda)\;)\;.
\end{equation}

Now, we will restrict our attention to qubit teleportation with non ideal
resource preparation $\kketp{S}{23}$, it is possible to give an
explicit expression for the minimum fidelity achieved by teleportation
on pure states.\\
If $\ket{\psi}$ is the original state, apart from a normalization
factor, the teleported state will be
\begin{equation}
\ket{\psi_\lambda}=U_\lambda S^TU_\lambda^\dagger\ket{\psi}\;,
\end{equation}
where $U_\lambda$ is the unitary operator related to the outcome
$\lambda$.\\ The minimum fidelity achieved by teleportation can be
written as follows
\begin{equation}
F_{min}=\min_{\ket{\psi}\in\hilb{}}\frac{|\bra{\psi}U_\lambda
S^TU_\lambda ^\dagger\ket{\psi}|^2} {\bra{\psi}U_\lambda S^*U_\lambda
^\dagger U_\lambda S^TU_\lambda ^\dagger\ket{\psi}}=
\min_{\ket{\psi}\in\hilb{}}\frac{|\bra{\psi}S^T\ket{\psi}|^2}
{\bra{\psi}S^*S^T\ket{\psi}}\;.
\label{min fidelity}
\end{equation}
Using a basis of $\hilb{2}\otimes\hilb{3}$ for which $\kketp{S}{23}$
is in the Schmidt-form, i.e. diagonal and positive, and noticing that
$F_{min}$ is independent of the normalization of $S$, we can choose
$S$ to be
\begin{equation}
S_ \epsilon =(1+\epsilon)\prj 0 +(1-\epsilon)\prj 1\;.
\label{S}
\end{equation}
The minimization can performed only on states $\ket{\psi (x)}$ of the form
\begin{equation}
\ket{\psi (x)}=\cos x\ket 0 + \sin x \ket 1\;,\qquad x\in[0,2\pi)\;,
\label{psi}
\end{equation}
because any phase would be irrelevant.
Substituting Eqs. (\ref{S}) and (\ref{psi}) in Eq. (\ref{min fidelity}) and
minimizing respect to $x$, one obtains
\begin{equation}
F_{min}=1-\epsilon^2\;.\label{fmin}
\end{equation}
With some little algebra, Eq. (\ref{fmin}) 
can be cast in a compact form independent
of basis and normalization as follows
\begin{equation}
F_{min}=4\frac{\det(\tilde S)}{Tr^2[\tilde S]}\;, 
\qquad \hbox{where} \quad\tilde S=\sqrt{S^\dagger S}\;.
\end{equation}
\section{Conclusions}
We studied the problem of characterizing Bell measurements, as
maximally entangled POVM's for measurements on composite systems. We
introduced operator-completeness relations and a simple matrix
approach to deal with bipartite systems. These allow us to write the
most general Bell-like POVM in a compact form. The role of spanning
sets of unitary operators has been emphasized, with attention to
unitary irreducible representations of groups. Bell observables
related to Bell POVM's have been explicitly derived. As direct
application of the matrix formalism, we evaluated the robustness of
teleportation to non maximality of the shared entangled resource and
to non ideality of the measurement.


\end{document}